# Capabilities of Deep Learning Models on Learning Physical Relationships: Case of Rainfall-Runoff Modeling with LSTM

Kazuki Yokoo[1], Kei Ishida[2,3], Ali Ercan[4], Tongbi Tu[5], and Takeyoshi Nagasato[1], Masato Kiyama[6], Motoki Amagasaki[6]

**Abstract**

This study investigates the relationships which deep learning methods can identify between the input and output data. As a case study, rainfall-runoff modeling in a snow-dominated watershed by means of a long- and short-term memory (LSTM) network is selected. Daily precipitation and mean air temperature were used as model input to estimate daily flow discharge. After model training and verification, two experimental simulations were conducted with hypothetical inputs instead of observed meteorological data to clarify the response of the trained model to the inputs. The first numerical experiment showed that even without input precipitation, the trained model generated flow discharge, particularly winter low flow and high flow during the snow-melting period. The effects of warmer and colder conditions on the flow discharge were also replicated by the trained model without precipitation. Additionally, the model reflected only 17–39% of the total precipitation mass during the snow accumulation period in the total annual flow discharge, revealing a strong lack of water mass conservation. The results of this study indicated that a deep learning method may not properly learn the explicit physical relationships between input and target variables, although they are still capable of maintaining strong goodness-of-fit results.

## 1. Introduction

Deep learning has recently started to be widely applied to issues in hydrology. There are already numerous applications as summarized in Shen [Citation error], Reichstein *et al*. (2019), Tahmasebi *et al*. (2020), and Xu and Liang (2021). Those applications have shown the capability of deep learning to generate outputs with high accuracy. Because its high capability, it is now expected that deep learning is interpretable in terms of physics (Shen, 2018; Reichstein *et al.*, 2019) although deep learning has been mostly used as a black-box model. Before that, however, there is a fundamental question: can deep learning really learn physical relations between variables properly?

1. Graduated School of Science and Technology, Kumamoto University
2. International Research Organization for Advanced Science and Technology, Kumamoto University
3. Center for Water Cycle, Marine Environment, and Disaster Management, Kumamoto University
4. Department of Civil and Environmental Engineering, University of California
5. School of Civil Engineering, Sun Yat-Sen University
6. Faculty of Advanced Science and Technology, Kumamoto University

This study utilizes long and short term memory (LSTM) network (Hochreiter and Schmidhuber, 1997; Gers *et al.*, 2000), which is a type of recurrent neural network (RNN), which is suitable for time series modeling. Especially, LSTM has capability to learn long term dependencies between input and target data. Because of such capability, there are many applications of LSTM, including the encoder-decoder form, in hydrology. For example, LSTM has been utilized for precipitation estimation and forecasting (Akbari Asanjan *et al.*, 2018; Misra *et al.*, 2018; Miao *et al.*, 2019; Tran Anh *et al.*, 2019), ground water modeling (Bowes *et al.*, 2019; Jeong and Park, 2019; Jeong *et al.*, 2020), soil water forecasting (Fang *et al.*, 2017; Fang and Shen, 2020), flood inundation forecasting (Hu *et al.*, 2019), and flow modeling and forecasting (Kratzert *et al.*, 2018; Tian *et al.*, 2018; Kratzert *et al.*, 2019b; Song *et al.*, 2019; Kao *et al.*, 2020; Liu *et al.*, 2020; Nearing *et al.*, 2020; Zhu *et al.*, 2020).

Some physical interpretations of deep learning have been conducted by some studies using LSTM (Kratzert *et al.*, 2018, 2019a). Kratzert *et al*.

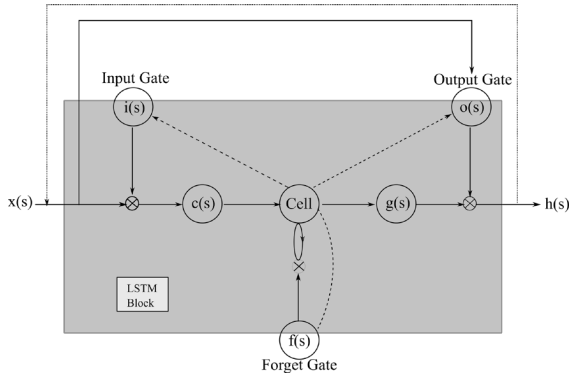

Figure 1. LSTM block.

(2018, 2019a) utilized LSTM for rainfall-runoff modeling in snow-dominated watersheds, and showed that an internal state (cell state) of the trained model behaves like the snow accumulation and melting process. In addition to that, Kratzert et al. (2019a) investigated correlations between the internal states and input variables. Thus, there have already been several studies on the interpretation of deep learning in terms of physics in hydrology. However, they did not investigate what kind of relations deep leaning actually learn in detail.

Contrarily, Karpatne et al. (2017a) claimed that physical relationship between variables is very complex, and thus the limited number of training data are frequently not sufficient for deep learning to properly grasp the true nature of the physical relationship. In addition, Jia et al. (2019) also claimed that deep learning learns simple relations between inputs and target data without basis of any physical laws, and thus it is possible to generate results that are inconsistent with physical laws. Meanwhile, they utilized LSTM to model water temperature in a lake and showed the inconsistency between the changes in air temperature obtained by LSTM and the summation of the incoming and outward energy fluxes. Deep learning model failed to learn the physical relationship in their study although they did not get into detail to investigate what kind of relations deep learning learns.

In this context, the study herein aims to address the issue of what kind of relations deep learning can grasp from the data. As a case study, rainfall-runoff modeling in a snow-dominated watershed is chosen. Flow discharge at a snow-dominated watershed is determined not only by precipitation, but by air temperature effects on hydrological processes involving snow accumulation and melting. There will be long-term dependencies between input and target variables. Hence, it would be suitable to apply LSTM. Daily precipitation and daily-mean air temperature data are employed as inputs, and daily flow discharge data are utilized as the target to train the rainfall-runoff model. After verifying the model accuracy, numerical experiments are conducted with hypothetical inputs to investigate relations the trained LSTM rainfall-runoff model learned.

## 2. Methods and Data

### 2.1. Long and short-term memory network

Daily atmospheric data, such as precipitation and air temperature, were employed as inputs, and outputs are daily flow discharge at a certain point of the study watershed. LSTM was implemented using a deep learning framework, PyTorch (Paszke et al., 2019). LSTM used in PyTorch generally maintains at least one memory cell, referred to as the LSTM block. An LSTM block consists of a memory and three gates (Figure 1). The memory is known as cell state $c$. When the LSTM block receives the input time series vector $x(s)$ at time $s$, it updates the cell state $c(s)$ and generates the hidden states $h(s)$. At the next time step $(s + 1)$, the hidden states $h(s)$ at time $s$ are replaced into the LSTM block together with the input time series vector $x(s + 1)$. The entire process can be expressed as follows (Eqs 1–5):

$$g_i(s) = \sigma(W_{ii}x(s) + b_{ii} + W_{hi}h(s-1) + b_{hi}) \quad (1)$$

$$g_f(s) = \sigma(W_{if}x(s) + b_{if} + W_{hf}h(s-1) + b_{hf}) \quad (2)$$

$$g_o(s) = \sigma(W_{io}x(s) + b_{io} + W_{ho}h(s-1) + b_{ho}) \quad (3)$$

$$c(s) = g_f(s) \otimes c(s-1) + g_i(s)$$
$$\otimes \tanh(W_{ic}x(s) + b_{ic} + W_{hc}h(s-1) + b_{hc}) \quad (4)$$

$$h(s) = g_o(s) \otimes \tanh(c(s)) \quad (5)$$

where $x$ is the input vector; $c$ and $h$ are the cell and hidden states, respectively; $g_i$, $g_f$, and $g_o$ are the input gate, forget gate, and output gate, respectively; $W$ and $b$ are the weights and biases, respectively; the two subscript letters for each weight and bias indicate the input/hidden state vector and gate, respectively, where the weights and biases are the parameters of the LSTM to be calibrated; $\sigma$ is the sigmoid function; and $\otimes$ indicates the Hadamard product. The LSTM block is recurrently utilized with an arbitrary length of the input time series vector, referred to as the input data length (IDL). When IDL is set to $T$, the LSTM block is repeatedly utilized $T$ times along the time $s = 1, 2, \cdots, T$. When $s$ reaches $T$, the hidden state is converted by the linear transformation to an output vector $y(t)$ at a certain time $t$. For the rainfall-runoff modeling in this study, when $s = T$, $s$ is equal to $t$; in other words, the last time step of the input time series $x$ was the same as the time step of the output vector $y$.

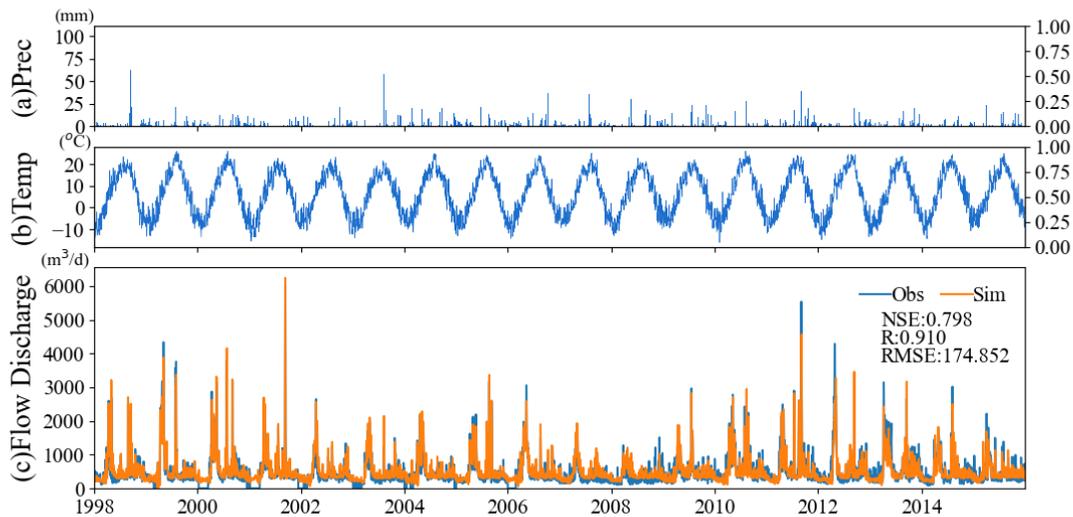

Figure 1. Observed (a) daily precipitation, mm; (b) daily-mean air temperature, °C; and (c) simulated and observed daily average flow discharge (m$^3$·s$^{-1}$). The right axes show the input normalized values.

## 2.2. Study Area and Data

The Ishikari River watershed (IRW), located in the Hokkaido region in the northern sector of Japan, was selected as the study watershed. The Ishikari River is the third longest river in Japan (268 km), and the IRW catchment area (14,330 km$^2$) is the second largest in in the country. This study investigated flow discharge at the Ishikari Ohashi (IO) gauging station, located 26.60 km from the river mouth of the Ishikari River at 43° 07' 20″ N, 141° 32' 32″ E. Flow discharge data from the IO station were obtained from the Water Information System (WIS) operated by the Ministry of Land, Infrastructure, Transport and Tourism, Japan (http://www1.river.go.jp/), providing hourly flow discharge for the period from 1998 to the present. Daily flow discharge was calculated by averaging hourly flow discharge data and using all available data on days when some parts were missing. In instances where all data were missing within a day, the day was removed from the LSTM training and statistical analyses.

Daily precipitation data were obtained from the Asian Precipitation–Highly Resolved Observational Data Integration Towards Evaluation of Water Resources (APHRODITE; Kamiguchi et al., 2010). APHRODITE is a daily gridded precipitation dataset, covering the entirety of Japan with a resolution of ~5 km from 1900–2015, and the daily basin-average precipitation was calculated for the IRW.

Daily mean air temperature data was provided by the Japan Meteorological Agency using the 45 meteorological stations located throughout the IRW. Analyses were performed over the entire coincident period of discharge and precipitation data, from 1998–2015. For deep learning applications, input data were normalized to a range between zero and one, and only the relative values were considered in the deep learning network. Furthermore, owing to the relatively small spatial variability, daily mean air temperature data were averaged over the 45 meteorological stations to obtain a representative daily mean air temperature for the IRW.

## 2.3. Model implementation

Similar to other deep learning methods, several hyper-parameter options for LSTM were utilized to train the model. Kratzert et al. (2018) set the input data length of LSTM network to 365 days for daily rainfall-runoff modeling to reflect the annual effects of snow accumulation and melting on model output. Each of the input data and the target data are normalized to the range from zero to one. This study also utilized 365 days of the input data length. Hidden state length was set to 50, and the gradient descent method using subsets (batches) of the training dataset was used to train the model. Batch size was set to 256, and the shuffling sampling method was used to generate each batch, randomly extracting samples from the dataset without replacement. Training with all data was considered an epoch. Adaptive moment estimation (Adam; Kingma and Ba, 2014) was utilized to optimize the learning rate of the gradient descent method. The mean square errors were used to calculate accuracy for the adjustment of the learning rate, and patience, defined as the number of epochs to check for overfitting of the validation dataset, was set to 30.

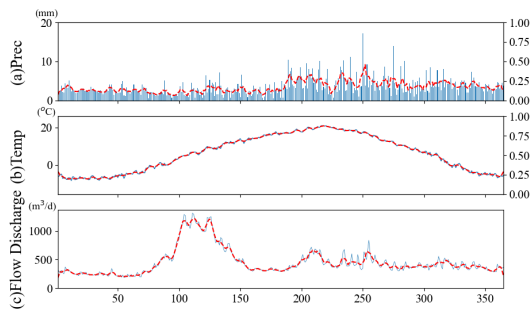

Figure 3. Observed daily (a) precipitation, mm; (b) air temperature, ºC; and (c) flow discharge, m³·s⁻¹ during the training period, with the five-day moving averages shown in red. The right axes show the input normalized values.

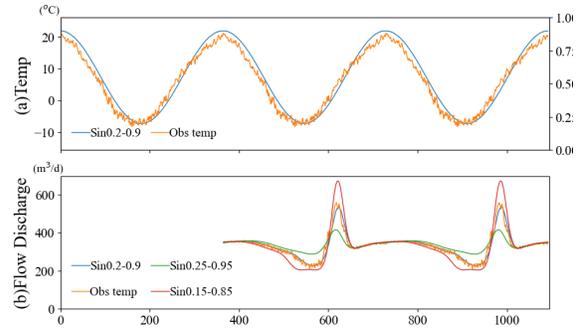

Figure 2. (a) Average daily-mean air temperature over the training period, and the normalized sinusoidal curve ranging between 0.2–0.9. The right axis shows the input normalized values. (b) Simulated flow discharge (m³·s⁻¹) generated by the trained model, with the daily-mean air temperature, and sinusoidal curves ranging between 0.15–0.8, 0.2–0.9, and 0.25–0.95.

As previously mentioned, the daily basin-average precipitation and the air temperature of the IRW were utilized as inputs, and the LSTM network model generated flow discharge data at a single location near the outlet of the watershed. The given dataset was separated into three groups: training (1998–2009), validation (2010–2012) to check for overfitting, and testing (2013–2015). The initial states of the learnable parameters (weights and biases) were generally given as random values for the training process to increase the robustness of the training process, although this may negatively affect the accuracy of the trained network; thus, this study ran the aforementioned training process 100 times. The network that displayed the greatest accuracy was selected as the final trained network and evaluated with the test dataset.

Several statistics were employed for accuracy assessment, including root mean square error (RMSE), correlation coefficient (R), and Nash-Sutcliffe efficiency (NSE). The RMSE, R, and NSE for the training (1998–2009), validation (2010–2012), and test periods (2013–2015) were: 144.7 m³·s⁻¹, 0.946, 0.877; 190.9 m³·s⁻¹, 0.940, 0.879; and 174.3 m³·s⁻¹, 0.904, and 0.800, respectively. Figure 2 displays that the model-simulated flow results were in good accordance with the observed data, further corroborated by the R values showing a high correlation between modeled and observed data. The NSE values indicate high accuracy for the training, validation, and test periods as well, indicating that the model is acceptable (NSE > 0.5; Moriasi *et al.*, 2007).

### 2.4. Numerical experiments

After model evaluation, two simulations were conducted to investigate the relationship between the input meteorological variables and the targeted flow discharge data learned by the LSTM for the IRW. For the numerical experiments, hypothetical time series variables were used as input to the trained model in order to evaluate the capability of deep learning to model physical processes.

The first numerical experiment (Case 1) was designed to assess the modeled response to temperature, and two hypothetical air temperature time series were evaluated with zero precipitation. Figure 3 shows the first set of experimental, average observed daily mean air temperatures values in the IRW during the training period. The normalized values ranged roughly from 0.2 (-7.3°C) to 0.9 (21.8°C), and the shape of the average daily air temperature throughout the year resembled a sinusoidal curve with a 365-day wavelength and an amplitude of 0.35. This sinusoidal curve was used as the second hypothetical input instead of direct air temperature in order to minimize fluctuations and clarify the relationship between the inputs and outputs of the model. In addition, the sinusoidal curve was artificially shifted by ± 0.05 (± 2.08°C) across the years analyzed, and assessed to investigate the effects on flow discharge (outputs).

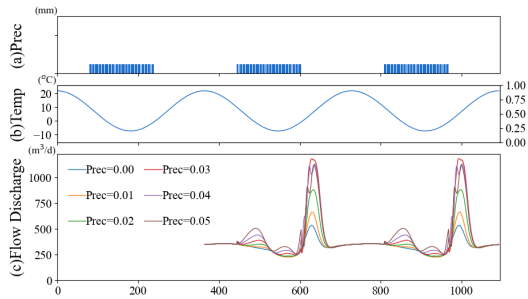

Figure 4. (a) Period of the given precipitation ranging between the normalized values of 0.0 to 0.05. (b) Sinusoidal daily air temperature (°C) ranging between the normalized values of 0.2–0.9. (c) Simulated flow discharge (m$^3$·s$^{-1}$) generated by the trained model with a sinusoidal air temperature and the constant precipitation during the snow-accumulation period.

The second numerical experiment (Case 2) used a hypothetical precipitation time series together with the sinusoidal air temperature defined above to investigate the effects of precipitation during the snow accumulation period on the model-simulated flow discharge during the snow melting period. For simplification, the hypothetical precipitation was held constant value during the snow accumulation period (mid-October to mid-March) for the study watershed. The average observed precipitation during the snow accumulation period ranged from 0.024 (2.37 mm) to 0.043 (4.33 mm) over the training period, with an average of 0.033 (3.31 mm); thus, the constant values were set from 0.01 (1.01 mm) to 0.05 (5.06 mm) at an increment of 0.01 (five total cases).

## 3. Results

### 3.1. Case 1: Zero precipitation

The LSTM network receives the given length (365 days) of input time series data, and generates daily flow discharge estimates. Therefore, the simulated flow discharge in Figure 4 was obtained by running the trained model with a 365-day slice of the input data every time step for two years. In this experiment, the model generated flow discharge through time using the average observed air temperatures and a zeroed precipitation. The resulting values varied with time, returning relatively low discharge values during the winter, rapidly increasing values with air temperature, and decreasing abruptly after reaching its peak. The peak value of the simulated flow discharge recorded was 558 m$^3$·s$^{-1}$ in the spring, while the average daily flow discharge observed over the training period was 1,100 m$^3$·s$^{-1}$, excluding peak flow events (Figure 3); thus, the modeled flow discharge during the snow-melting period was slightly greater than half of the observed values when holding precipitation constant at zero.

The sinusoidal curve between 0.2–0.9 replicated the average observed air temperature (Figure 4a), and the simulated flow discharge using this data behaved similarly to that of the averaged observed air temperature. Based on this sinusoidal curve, the air temperature was shifted by ± 0.05 (± 2.08°C) over the year for further investigation, and it was found that the increase in air temperature generated increased flow discharge during the winter, and decreased levels during the snow melting period. Conversely, the sinusoidal curve under decreased temperatures revealed decreased flow discharge during the winter, and increased levels during the snow melting period.

### 3.2. Case 2: Non-zero precipitation during snow accumulation period

In Case 2, constant precipitation during the snow accumulation period was used for model input, together with the above sinusoidal air temperature (Figure 5). The given precipitation depth was changed from 0.00 to 0.05 at increments of 0.01 (1.01 mm). The results showed that the inclusion of precipitation increased the flow discharge not only during the periods of snow-accumulation, but during the snow-melting as well. When the given precipitation was 0.03, nearly equal to the average precipitation during the snow accumulation period over the training period, the peak value of the simulated flow discharge during the snow-melting period, 1,184.6 m$^3$·s$^{-1}$, was similar to the average daily flow discharge of the training period, 1,100 m$^3$·s$^{-1}$, excluding peak flow events (Figure 3c), indicating that the trained model properly reproduced the average behavior of snowmelt flow discharge.

Simulated flow discharge increased during the snow accumulation period as higher precipitation levels were implemented.; however, a similar increase was not observed during the snow-melting period across all precipitation levels. The peak flow discharge during the snow-melting period gradually increased with precipitation, from 0.01–0.03, with peak flow discharge values of 663.1 $m^3 \cdot s^{-1}$, 881.3 $m^3 \cdot s^{-1}$, and 1,184.6 $m^3 \cdot s^{-1}$ recorded at 0.01, 0.02, and 0.03, respectively; however, the increase in constant precipitation from 0.03 to 0.04 reduced flow discharge by 55.2 $m^3 \cdot s^{-1}$ during the snow-melting period. A further decrease was observed with an increase in the constant precipitation from 0.04 and 0.05, although the peak values were similar. The average precipitation observed values during the snow accumulation period ranged from 0.033 to 0.043, and the revealed decrease in simulated flow discharge due to the increase in precipitation during the snow accumulation period was not corroborated but the observations.

The overall increase in volume of flow discharge over the year was small compared to the changes in precipitation volume. When increasing daily precipitation from 0.01 (1.01 mm) to 0.05 (5.06 mm) during the 160 day snow-accumulation period, the simulated flow discharge increased by 0.35 $km^3$, 1.12 $km^3$, 2.43 $km^3$, 2.67 $km^3$, and 2.83 $km^3$, respectively, for the year when compared to the zero precipitation simulation. The size of catchment area upstream of the Ishikari Ohashi gauging station is approximately 12,850 $km^2$. The annual total volumes of the given precipitation from 0.01 to 0.05 are 2.08 $km^3$, 4.16 $km^3$, 6.24 $km^3$, 8.32 $km^3$, and 10.40 $km^3$, respectively. Thus, the modeled results reflected only 17–39% of the annual total flow discharge output for a given precipitation volume.

## 4. Discussion

Case 1 demonstrated that even without precipitation, LSTM still generated some levels of base flow discharge (Figure 4). It was also able to replicate, to some degree, the observed reduction in flow discharge during the winter, and the increase during the snow-melting period. Furthermore, it captured the effects of warmer and colder years on flow discharge without precipitation. Thus, the results in Case 1 indicate that LSTM directly learned the relationship between air temperature and flow discharge to some extent without consideration of interaction with precipitation.

Case 2, alternatively, showed that the model did not properly reproduce the flow discharge during the snow-melting period, even when the given precipitation during the snow accumulation was within the observed range. The model predicted decreased flow discharge during the snow-melting period, with precipitation levels > 0.03 (3.03 mm) during the snow accumulation period. Additionally, the trained model reflected only 17–39% of the flow discharge output for a given level of precipitation. We also calculated the ratio of flow discharge to precipitation for the training period (1998-2009), which is approximately 87%. Compared to this percentage, the ratios, obtained from the simulated results in Case 2, were unrealistically small. Obviously, water mass was not conserved between the input and output variables. As described in Introduction section above, Jia *et al.* (2019) showed that LSTM failed to conserve the energy. The target data of their study was not energy, but water temperature. Contrarily, this study targeted flow discharge. Nevertheless, the mass conservation was not held. In other words, LSTM did not learn the mass relations between precipitation and flow discharge. The training dataset for deep learning need to be normalized. Mass is not conserved between the input and the targe data because fundamentally the required information of the mass conservation is not given to deep learning.

Karpatne *et al.* (2017b) and Jia *et al.* (2019) developed an approach to remediate the inconsistency in the energy conservation for water temperature modeling by deep learning by adding physical constraints to the loss function for the training process. Their approach can reduce the errors in some other variables as well as the target variables during the training process. This approach may also be able to provide some improvements on the issues described above in this study. Meanwhile, there is no guarantee that their approach enables LSTM to learn physical relations.

Finally, the simulations in this study were conducted using a single-trained model at a study watershed. The results may have been affected by the model, its hyperparameters, input variables, or the study watershed itself; however, the model trained in this study showed strong statistical values. LSTM networks are increasingly being applied to various issues in hydrology. Although deep learning may be a powerful tool for revealing complex relationships between input and target variables, the results of the study here indicate that a deep learning method, although producing accurate results, may fail to properly capture variable relationships.

## 5. Conclusions

This study investigated what kind of relations between the input and target variables LSTM can grasp. As a case study, rainfall-runoff modeling in a snow-dominated watershed was selected. The trained model was utilized for two numerical experiments to explore the ability of LSTM to learn the physical relationship between the input meteorological data (precipitation and temperature), and the target flow discharge data. Although the trained model showed high accuracy in terms of the goodness-of-fit test, the experimental results indicated that the trained model could not properly capture the relationship between input and target data.

The results of this study here indicate that modelers and decision makers should be careful when employing deep learning methods, as it may reveal unexpected and unintended relationships between the given data. Thus, deep learning methods should be cautiously used after sufficient justification when applied to learn new physical relationships.